\documentclass[12pt]{article}
\usepackage{amsfonts,epsf}

\def\d{{\rm d}}
\def\mi{{\rm i}}

\def\G{\mathop{\Gamma}\nolimits}
\def\Re{\mathop{\rm Re\,}\nolimits}

\def\e{\mathop{\rm e}\nolimits}

\def\hf{{\textstyle{1 \over 2}}}
\def\qt{{\textstyle{1 \over 4}}}
\def\N2{{\textstyle{N \over 2}}}

\def\defi{\stackrel{\rm def}{=}}
\def\si{\!\!\! &}
\def\se{& \!\!\!}

\newcommand{\beq}{\begin{equation}}
\newcommand{\eeq}{\end{equation}}
\newcommand{\bea}{\begin{eqnarray}}
\newcommand{\eea}{\end{eqnarray}}

\title{The general 1D Schr\"odinger equation as an exactly solvable problem}

\author{{\bf Andr\'e Voros}\footnote{Also at: 
Institut de Math\'ematiques de Jussieu--Chevaleret (CNRS UMR 7586), 
Universit\'e Paris 7, F-75251 Paris CEDEX 05, France.}\\
\\
CEA, Service de Physique Th\'eorique de Saclay\\
(CNRS URA 2306)\\
F-91191 Gif-sur-Yvette CEDEX, France\\
E-mail : {\tt voros@spht.saclay.cea.fr} }

\begin{document}
\maketitle

We review an exact WKB resolution method
for the stationary 1D Schr\"o\-dinger equation 
with a general polynomial potential.
This contribution covers already published material:
we supply a commented summary here, stressing
a few aspects which were less highlighted before.
As some of our earlier papers needed later corrections,
we also recapitulate these here in footnotes in the bibliography.
The latter, not meant to be exhaustive, focuses on narrowly relevant works
plus some broader ones that do offer more extensive bibliographies.

\section{General setting}

We will treat a Schr\"odinger equation on the real line
with a polynomial potential function $V$ (real, normalized as
$V(q)= {+ \, q^N + \, v_1 \,q^{N-1} + \cdots + \, v_{N-1} \,q} $):
\beq
\label{SSE}
\Bigl( \,-{\d^2 \over \d q^2} + \bigl[ V(q)+\lambda \bigr] \ \Bigr)
\psi (q)=0 ,
\eeq
(a  general Sturm--Liouville problem).
It is convenient to pose the problem on a {\sl half-line\/},
initially $q \in [0,+\infty)$;
exact treatments however rely on analytical continuations
to the {\sl whole complex domain\/}: 
$q \in {\mathbb C}$, $(\vec v, \lambda) \in {\mathbb C}^N$
(where $\vec v \equiv (v_1,\ldots,v_{N-1})$).

\subsection{Outline of results \cite{V5}--\cite{V7}}

The traditional view about eq.(\ref{SSE}) (in any dimension) is that

\noindent - $N=2$ is exactly solvable: the harmonic oscillator,
``the nice" case;

\noindent - no other $N$ is solvable (generically):

$\bullet \ N=1$ gives the {\bf Airy} equation
(more transcendental than the higher-degree $N=2$, a paradox!)

$\bullet \ N \ge 3$ yield the {\bf anharmonic oscillators},
the ground of choice for all approximation methods
(perturbative, semiclassical \ldots).
\medskip

We see eq.(\ref{SSE}) instead as an {\bf exactly solvable problem for any $N$},
thanks to an {\sl exact WKB\/} method with two key ingredients:

\noindent - a semiclassical analysis exploiting {\bf zeta-regularization} 
\cite{VC,JL};

\noindent - exact quantization conditions that are Bohr--Sommerfeld-like but
{\bf selfconsistent}, and akin to the
{\bf Bethe Ansatz} equations that solve many exact models 
in 2D statistical mechanics \cite{DT,SU} or quantum field theory \cite{BLZ}.

Moreover, degrees $N$ and $4/N$ behave similarly in this framework
(a qualitative duality) \cite{V4}:

$\bullet \ N=1$ is transcendental like $N=4$, clarifying the above paradox;

$\bullet \ N=2$ (selfdual) becomes
completely explicitly solvable as a degenerate, rather pathological, case.
A {\sl countable family\/} of similarly solvable cases is actually obtained
\cite[Sec.4]{V7}:
all the generalized eigenvalue problems
${ [ -{\d^2 \over \d q^2} + (q^N \!+\! \Lambda \, q^{{N \over 2} -1}) ]
\, \psi (q) = 0 }$ (for even degrees $N>0$).

\subsection{A few early sources}

{\sl Exact asymptotics\/} (which uses drastically divergent representations 
to perform fully exact calculations) is a paradoxical idea; 
its advent rests on various seminal developments which mainly
took place in the '70s (restricting to works we felt as directly influential; 
a comprehensive bibliography would need much more space):

\noindent - Balian and Bloch's representation of quantum mechanics,
where the latter can be {\sl exactly rebuilt in principle\/} 
just from the classical trajectories, provided these are taken
in their full {\sl complex\/} extension \cite{BB};

\noindent - the {\sl exact geometrical\/} content of high-frequency asymptotics:
singularities, wave front sets, and their propagation properties \cite{H1,C,DG};
(reviewed in \cite[chap.~VIII]{H2})

\noindent - the same in the {\sl analytic (not $C^\infty$)\/} category
(a {\sl key step\/} to exactness): ramified Cauchy problem \cite{L},
analytic pseudo-differential operators \cite{BMK}, 
hyper- and micro-functions \cite{SKK}, localized Fourier transforms \cite{BI};
(reviewed in \cite{SJ})

\noindent - {\sl Borel resummations\/} of asymptotic series \cite{DI}, 
large-order perturbation theory \cite{BW2}, tunneling (e.g., \cite{H,BPV,M}),
instantons and Zinn-Justin's conjectures \cite{ZJ};

\noindent - seemingly unrelated direct monodromy calculations resulting in
{\sl exact functional relations\/} for Stokes multipliers \cite{S}.

The Balian--Bloch idea led to a first concrete exact WKB algorithm 
for 1D problems like eq.(\ref{SSE}),
based on (generalized) Borel resummations \cite{V1,KT,DDP,DP,BM,HKT},
and best formalized in \'Ecalle's framework of {\sl resurgent functions\/} 
\cite{E}; the explicit exact WKB results are indeed special instances of 
{\sl bridge equations\/} in resurgence algebras (or resurgence equations).
Still, in this approach, one basic resurgence theorem is not yet fully proved
\cite[thm 1.2.1 and its Comment]{DP},
and a fairly explicit solution stage has only been reached for 
homogeneous polynomial potentials, rather tortuously \cite{V2}--\cite{V4}.
Otherwise, 
many developments have been achieved in this and neighboring directions:
proofs of Bender--Wu \cite{KT,DP2} and Zinn-Justin conjectures \cite{DDP,DP},
hyperasymptotics (reviewed in \cite[Part II]{HKT}), 
results on Painlev\'e functions (reviewed in \cite[Part III]{HKT}).

This presentation is about a more direct exact WKB method \cite{V5}--\cite{V7}
(surveyed in \cite{V8}), based upon functional relations 
{\sl \`a la\/} Sibuya \cite{S}, plus crucial inputs of
{\sl zeta-regularization\/} both at the quantum and classical levels.
At its core, it needs just one exact Wronskian identity plus
a structure equation for a pair of spectral determinants,
to generate the exact solution as a fixed point of an explicit system
of nonlinear equations (``exact quantization conditions", but also
{\sl Bethe-Ansatz\/} equations in an integrable-model interpretation \cite{DT}).
The Wronskian identity precisely subsumes the resurgence relations found
in the Borel-transform approach, whereas the newer exact quantization conditions
(which achieve the solving task) have no clear counterpart in the Borel plane
as far as we know.
This direct approach also makes it easier to treat the general case at once,
and it is now fully proven for homogeneous potentials \cite{A}.

\section{Basic ingredients}

\subsection{The conjugate equations \cite{S}}

Just as a polynomial equation is better handled by treating 
all its conjugate roots together, the original differential equation
(\ref{SSE}) is to be supplemented by its {\sl conjugates\/}, 
defined using
\beq
\label{CEQ}
V^{[\ell]}(q) \defi \e^{-\mi\ell\varphi} V(\e^{-\mi\ell\varphi/2}q), 
\quad \lambda^{[\ell]} \defi \e^{-\mi\ell\varphi} \lambda , \qquad
\mbox{with } \varphi \defi {4\pi \over N+2} \ ;
\eeq
there are $L$ distinct conjugate equations labeled by 
$\ell \in {\mathbb Z} / L {\mathbb Z}$, where
\beq
L = \Biggl\{ { N+2 \quad \mbox{generically} \hfill \atop
\N2 + 1 \quad \mbox{for an \it even polynomial }V(q) .} \Biggr.
\eeq

\subsection{Semiclassical asymptotics}

\qquad $\bullet$ WKB formulae for solutions of the Schr\"odinger equation
\cite{S}:
a solution of eq.(\ref{SSE}) which is {\sl recessive\/} (= decaying)
for $q \to +\infty$ is specified by the WKB behavior
\beq
\label{WKB}
\psi_\lambda (q) \sim \Pi_\lambda(q)^{-1/2} 
\, \exp \ -\!\!\int^q_Q \Pi_\lambda(q') \d q' \, , \qquad
\Pi_\lambda(q) \!\defi\! (V(q) \!+\! \lambda)^{1/2} ,
\eeq
where $\Pi_\lambda(q)$ is the classical forbidden-region momentum
($\Pi_\lambda(q) \, \d q $ is the Agmon metric).
Reexpansions in descending powers of $q \to +\infty$ yield
\bea
[V(q) +  \lambda]^{-s+ 1/2} \si \sim \se
\sum_\sigma \beta_\sigma (s;\vec v,\lambda) \, q^{\,\sigma-Ns} \quad
(\sigma=\N2,\ \N2 \!-\! 1,\ \N2 \!-\! 2,\ \ldots) , \\
\label{ASQ}
\psi_\lambda (q) \si \sim \se
\e^{\,\mathcal C} q^{-N/4 \, -\beta_{-1}(\vec v)}
\exp \, \Biggl\{ -\!\!
{\displaystyle \sum_{\{\sigma>0\}}} \!\beta_{\sigma-1}(\vec v)
\,{\textstyle q^{\sigma} \over \textstyle \sigma} \Biggr\}
\eea
where $\beta_{\sigma-1} (\vec v) \defi \beta_{\sigma-1} (0;\vec v,\lambda)$
(independent of $\lambda$ for $\sigma \ge 0$ when $N > 2$).

Another solution, recessive for $q \to + \e^{-\mi\varphi/2} \!\infty$,
arises through the {\sl first conjugate\/} equation:
\beq
\Psi_\lambda (q) \defi \psi_{\lambda^{[1]}}^{[1]} (\e^{\mi\varphi/2}q) .
\eeq
As the asymptotic form (\ref{ASQ}) and its analogs for $\Psi_\lambda (q)$ 
and their $q$-derivatives have overlapping sectors of validity,
this {\bf explicit exact Wronskian} follows:
\beq
\label{EWR}
\Psi_\lambda (q)\psi'_\lambda (q)-\Psi'_\lambda (q)\psi_\lambda (q)
\equiv \e^{{\,\mathcal C}+ {\mathcal C}^{[1]}}
2\,\mi\e^{\mi\varphi/4} \e^{\mi\varphi\beta_{-1}(\vec v)/2} \quad (\ne 0).
\eeq
\smallskip

$\bullet$ Spectral asymptotics (with $E \equiv - \lambda$):
the Schr\"odinger operator over the whole real line with
the confining potential $V(|q|)$ has a purely discrete spectrum
$\{ E_k \}_{k \in \mathbb N}$ ($E_k \uparrow +\infty$).
Due to {\sl parity-splitting\/},
${\mathcal E}^+ \defi \{ E_{2k} \}$ is the Neumann,
and ${\mathcal E}^- \defi \{ E_{2k+1} \}$ the Dirichlet, spectrum 
(the boundary conditions being at $q=0$).
As a corollary of the WKB formula (\ref{WKB}) on eigenfunctions,
the eigenvalues obey the {\sl semiclassical Bohr--Sommerfeld\/} condition
\beq
\label {BSC}
S(E_k) \sim k + \hf \qquad \mbox{for integer } k \to + \infty ,
\eeq
in terms of the {\sl classical action function\/}
$ S(E) \defi \oint_{\! \{ p^2 + V \! (q)=E \} } \!
 {p \,\d q \over 2 \pi}$,
which behaves as $ b_\mu E^\mu$ for $E \to +\infty$
with $\mu = \hf +{1 \over N}$ (the {\sl order\/}).

Further expansion of eq.(\ref{BSC}) to all orders in $k ^{-1}$
followed by reexpansion
in descending powers of $E$ results in a complete asymptotic
($E \to + \infty$) eigenvalue formula, of the form \cite{V5}
\beq
\label{SCE}
\sum_\alpha  b_\alpha {E_k}^\alpha \sim k + \hf \quad
\mbox{for integer } k \to + \infty \quad \textstyle 
( \alpha= \mu,\ \mu-{1 \over N},\ \mu-{2 \over N}, \ldots )
\eeq
(the $b_\alpha$ are polynomial in the $\{v_j\}_{j \le (\mu-\alpha)N}$,
and also parity-dependent but not above $\alpha=-3/2$
nor if $V$ is an even polynomial).
\smallskip

\subsection{Spectral functions \cite{VC}}

\qquad $\bullet$ Spectral {\bf zeta functions} (parity-split for later use):
\beq
\label{SQZ}
Z^\pm(s,\lambda) \defi \! \sum_{k \ {\rm even \atop odd}}
(E_k + \lambda) ^{-s} \quad (\Re s > \mu), \quad
(\mbox{and } Z(s)=Z^+(s)+Z^-(s))
\eeq
extend to meromorphic functions of $s \in \mathbb C$, 
{\sl regular at} $s=0$, all due to eq.(\ref{SCE}).
\smallskip

$\bullet$ Spectral {\bf determinants}: {\sl formally\/} meant to be
\beq
\label{FQD}
D^\pm(\lambda ) = \mbox{``} \!\prod\limits_{k \ {\rm even \atop odd}} 
(\lambda \!+\! E_k) \mbox{ "} ,
\eeq
they get rigorously specified through
\beq
\label{SQD}
\log D^\pm(\lambda ) \defi - \partial_s Z^\pm(s,\lambda)_{s=0}
\qquad  (\mbox{\it zeta-regularization}) ,
\eeq
and further evaluated by {\sl limit formulae\/} which we call
{\sl structure equations\/} (they are like Hadamard product formulae, 
but with {\bf no undetermined constants whatsoever}, a crucial property here):
\bea
\label{EML}
\log D^\pm(\lambda ) \si=\se
\lim_{K \to + \infty}  \Biggl\{ \sum_ {k<K} \log (E_{k} + \lambda)
+ \hf \, \log (E_{K} \!+\! \lambda) \Biggr. \nonumber \\
\textstyle (k,\, K \ {\rm even \atop odd}) \si\se \\[-10pt]
\si\se \mbox{\it (counterterms:)} \quad - \hf \sum_{ \{\alpha >0 \} }
\Biggl. \vphantom{\sum_K}
 b_{\alpha} (E_K)^\alpha \Bigl[ \log E_K - \textstyle {1 \over \alpha} \Bigr]
\Biggr\} . \nonumber
\eea

As corollaries for $D$ (and $D^\pm$):

\noindent - $D$ is an {\bf entire} function in $(\lambda, \vec v)$,
of order $\mu$ in $\lambda$;

\noindent - $\log D$ has a large-$\lambda$ expansion
of a severely constrained  (``standard" or ``canonical") form, 
or {\sl generalized Stirling expansion\/} \cite{JL}:
\bea
\label{GSE}
\log D(\lambda) \si\sim\se \Bigl[ a_1 \lambda (\log \lambda -1) \Bigr]
\, + a_0 \log \lambda
\quad \Biggl[ \ + \sum_{ \mu_k \notin {\mathbb N} }
a_{\mu_k} \lambda^{\mu_k} \Biggr] \\
&& \quad \mbox{(degree 1) \quad (degree 0) \quad (degrees
$\mu_k \downarrow -\infty$)} \nonumber
\eea
where obviously no term can occur with degree $>\mu$
(growth restriction; e.g., $a_1 \equiv 0$ as soon as $N>2$);
but above all,

\noindent any extraneous {\bf power terms} 
$b_n \lambda^n \ (n \in {\mathbb N})$,
including {\bf additive constants} ($b_0 \lambda^0$), are {\bf banned} outright
(canonical constraint).

For $N=2$, eqs.(\ref{EML}), resp. (\ref{GSE}), 
essentially restore classic results:
the analytical continuation by the {\sl Euler--Maclaurin formula\/}, 
resp. the {\sl Stirling expansion\/}, for
$ -\partial_s \zeta(s,\lambda)_{s=0} \equiv
\log \,  \sqrt{2\pi} / \Gamma(\lambda) $ (Lerch formula),
with $\zeta(s,\lambda)=$ the Hurwitz zeta function.
Those two founding formulae of all asymptotics
involve zeta-regularization, as the Lerch formula shows;
by inference, WKB theory (a branch of asymptotics) also ought 
to be intimately tied to zeta-regularization,
but {\sl this\/} aspect traditionally goes {\sl unnoticed\/}.

\section{Fundamental exact identities}

\subsection{Canonical normalization of recessive solution}

The solution $\psi_\lambda (q)$ described by eqs.(\ref{WKB}), (\ref{ASQ})
to be recessive for $q \to +\infty$, still awaits normalization.
It proves useful to fix it also at $q=+\infty$, as
\beq
\label{CNC}
\psi_\lambda (q) \sim \Pi_\lambda(q)^{-1/2}
\, \exp \int_q^{+\infty} \Pi_\lambda(q') \,\d q' \qquad (q \to +\infty) .
\eeq
However, $\int_q^{+\infty} \Pi_\lambda(q') \,\d q'$
(``the Agmon distance from $q$ to $+\infty$") diverges:
we then wish to define this ``improper action integral" as
\beq
\label{PRE}
\Biggl[ \int_q^{+\infty} (V(q') + \lambda)^{1/2 \, -s}
\,\d q' \Biggr]_{s \to 0}
\eeq
(analytical continuation from $\{ \Re s > \mu \}$ to $s=0$),
but $s=0$ is a {\sl singular point\/} in general: a simple pole
of residue $ N^{-1} \beta_{-1}(\vec v) $ (cf. eq.(\ref{ASQ})). Now:

$\bullet \ \beta_{-1}(s;\vec v) =0$ {\sl quite frequently\/}
(for any polynomial $V(q)$ of odd degree $N$; 
also for any even polynomial of degree $N=4M$;
for any $V(q)=q^N,\ N \ne 2$); then $s=0$ becomes a regular point
and the prescription (\ref{PRE}) is fine;

$\bullet$ when $\beta_{-1}(s;\vec v)\ne 0$, a stronger regularization
is suggested by the quantum--classical correspondence.
``Classical spectral functions" are naturally definable,
by mimicking the quantum formulae (\ref{SQZ}), (\ref{SQD}), as
\bea
Z_{\rm cl}(s,\lambda ) \si\defi\se 
\int_{{\mathbb R}^2} \! {\d q \,\d p \over 2\pi} \, 
( p^2+V(|q|) + \lambda ) ^{-s} \\
\log D_{\rm cl}(\lambda ) \si\defi\se -\partial_s Z_{\rm cl}(s,\lambda)_{s=0} .
\eea
These formulae together imply a classical counterpart to 
the formal eq.(\ref{FQD}),
\beq
\label{FCD}
\mbox{``} \textstyle \int_{-\infty}^{+\infty}
(V(|q|) + \lambda)^{1/2} \,\d q \ " = \log D_{\rm cl} (\lambda ) ;
\eeq
using parity symmetry, this further splits to give the pair \cite[Sec.1.2.2]{V7}
\beq
\label{CPS}
D_{\rm cl}^\pm(\lambda ) = \Pi_\lambda(0)^{\pm 1/2} 
\exp \, \mbox{``} \textstyle \int_0^{+\infty} \Pi_\lambda(q) \,\d q \ " ;
\eeq
$\log D_{\rm cl}^\pm(\lambda )$ are also the divergent parts of 
the large-$\lambda$ expansions (\ref{GSE}) for $\log D^\pm(\lambda )$, 
so they are canonical as well.
The idea is then to {\sl define\/} $\int_0^{+\infty} \Pi_\lambda(q) \,\d q$ 
to be $\hf \log D_{\rm cl} (\lambda )$
(and $\int_q^{+\infty}=\int_0^{+\infty}-\int_0^q$).
This in turn fixes the recessive solution, with a normalization 
$\e^{\mathcal C} \not\equiv 1$ in eq.(\ref{ASQ})
(an {\sl anomaly\/} factor, which would turn up elsewhere if not here):
\beq
\label{ANC}
{\mathcal C} \equiv \textstyle {1 \over N} 
[ - (2 \log 2) \beta_{-1}(s ; \vec v)
+ \partial_s ( {\beta_{-1}(s ; \vec v) \over 1-2s} ) ] _{s=0} \ .
\eeq

As examples of the resulting improper action integrals: \cite{V9}
\beq
\matrix{ \hfill \int_0^{+\infty} (q^N + \lambda)^{1/2} \,\d q \si=\se
- \hf \, \pi^{-1/2} \G (\hf + \mu) \G (-\mu) \, \lambda^\mu \quad 
(N \ne 2; \ \mu  = \hf \!+\! {1 \over N})  \cr
\hfill \int_0^{+\infty} (q^2 + \lambda)^{1/2} \,\d q \si=\se
-\qt \, \lambda (\log \lambda -1) \hfill
(\mbox{anomalous when } \lambda \ne 0) \cr
\int_0^{+\infty} (q^4 + v q^2)^{1/2} \,\d q
\si=\se  - {1 \over 3} \, v^{3/2} . \hfill }
\eeq

Remarks: 

\noindent - a ``classical determinant" $D_{\rm cl}$ (or $D_{\rm cl}^\pm$)
is {\sl not\/} an entire function of $\lambda$ like $D$:
the discrete spectrum of zeros of $D$ classically becomes 
a continuous branch cut for $D_{\rm cl}$.

\noindent - our treatment of the normalization for ${\beta_{-1} \not\equiv 0}$ 
was initially flawed \cite{V5,V6}:
we erroneously prejudged the natural answer to remain ${\mathcal C}=0$,
so we had temporary inconsistencies of normalization in the anomalous case;
these affect eqs.(15--19), (26) in \cite{V5} and Secs.1.1, 2.1, 2.2 in \cite{V7}, 
but cancel themselves spontaneously thereafter; 
the error is repaired from \cite{VE} onwards.

\subsection{Basic exact identities}

Under the canonical normalization for the recessive solution,
and with  $' = {\d \over \d q} $, 
\beq
\label{BID}
D^+(\lambda) \equiv - \psi'_\lambda(0), \qquad resp. \qquad
D^-(\lambda) \equiv \psi_\lambda(0).
\eeq
The idea, assuming $N>2$, is first to prove the logarithmic
$\lambda$-derivatives of eqs.(\ref{BID})
by analytically computing traces of the resolvent kernel,
and to $\lambda$-integrate back:
then, no integration constants can reappear precisely because 
all four logarithms have the canonical large-$\lambda$ behavior (\ref{GSE}),
in particular
\beq
\psi'_\lambda(0) \sim - D_{\rm cl}^+(\lambda), \quad resp. \quad
\psi_\lambda(0) \sim D_{\rm cl}^-(\lambda)
\eeq
(from eqs.(\ref{CPS}) and (\ref{CNC}) at $q=0$, 
{\sl also valid for\/} $\lambda \to +\infty$). 
In depth, eqs.(\ref{BID}) arose by controlling
{\sl all integration constants\/} without exception, from eq.(\ref{EML}) onwards.

The basic exact identities (\ref{BID}) can now be substituted
into the explicitly known Wronskian (\ref{EWR}) computed at $q=0$, resulting in
a {\sl functional relation between spectral determinants\/}
or ``Wronskian identity", also fundamental:
\beq
\label{FFR}
\e^{+\mi\varphi/4} D^{[1]+} (\e^{-\mi\varphi} \lambda ) D^-(\lambda )
-\e^{-\mi\varphi/4} D^+(\lambda ) D^{[1]-} (\e^{-\mi\varphi} \lambda ) 
\equiv 2 \,\mi \e^{+\mi\varphi\beta_{-1}(\vec v) /2} .
\eeq

\section{Exact quantization conditions}

At first glance, the exact functional relations (\ref{FFR}) look 
grossly underdetermined; like the original Wronskian formula (\ref{EWR}), 
they amount to one equation for two unknown functions.
The exact information here seems to present a ``missing link" throughout.
This is however a delusion, because those unknown functions have a truly 
special form here, embodied in the structure equations (\ref{EML})
(themselves consequences of zeta-regularization),
and this precisely fills that information gap.

\subsection{Degenerate cases}

\qquad $\bullet \ N=2: \quad
\bigl[ -{\d^2 \over \d q^2} + (q^2+ \lambda) \bigr] \, \psi (q) = 0$
\qquad (harmonic oscillator).

This is the most singular case: $\beta_{-1}=\lambda/2 \not\equiv 0$,
and it has $\lambda$-dependence.
Now, $\varphi=\pi$ makes eq.(\ref{FFR}) degenerate: for $\lambda$ real, 
it separates into real and imaginary parts, which recombine as 
$ D^+(\lambda) D^-(-\lambda) = 2 \,\cos \, {\pi \over 4}(\lambda \!-\! 1) $.
The zeros are then obvious, and they clearly split between $D^+(\lambda)$ 
or $D^-(-\lambda)$ according to their sign; thereupon, 
the structure equations (\ref{EML}) yield the explicit solutions
\beq
D^+(\lambda) = {2^{-\lambda/2} \, 2\sqrt\pi \over \G ( {1+\lambda \over 4} )} , 
\qquad
D^-(\lambda) = {2^{-\lambda/2}\sqrt\pi \over \G ( {3+\lambda \over 4} )} .
\eeq

$\bullet \ N$ even: the generalized eigenvalue problems
${\bigl[ -{\d^2 \over \d q^2} + 
(q^N \!+\! \Lambda \, q^{{N \over 2} -1}) \bigr] \, \psi (q) = 0}$
are also exactly solvable \cite[Sec.4]{V7}:
setting $\nu={1 \over N+2}$, eqs.(\ref{FFR}) and (\ref{EML}) likewise lead to 
the relevant Neumann and Dirichlet determinants for ${q \in [0,+\infty)}$, as
\beq
D_N^+(\Lambda ) =
-{ 2^{-\Lambda /N} (4\nu)^{\nu(\Lambda +1) +1/2} \G(-2\nu)
\over \G (\nu(\Lambda -1) + 1/2) } \, , \quad 
D_N^-(\Lambda ) =
{ 2^{-\Lambda /N} (4\nu)^{\nu(\Lambda -1) +1/2} \G(2\nu)
\over \G (\nu(\Lambda +1) + 1/2) } \, ;
\eeq
whereas over the whole real $q$-axis,
\beq
\det \, [ \textstyle -{\d^2 \over \d q^2} 
+ (q^N \!+\! \Lambda \, q^{{N \over 2} -1}) ] =
\left\{ \matrix{ D_N^+(\Lambda ) D_N^-(\Lambda ) \hfill &&
{\rm if\ } N \equiv 2 \pmod 4 \hfill \cr
(\sin \, \pi \nu)^{-1}  \,\cos \, \pi \nu \Lambda && 
{\rm if\ } N \equiv 0 \pmod 4 .}
\right.
\eeq
Remark: actually, the potentials $(q^N \!+\! \Lambda \, q^{{N \over 2} -1})$
are {\sl supersymmetric\/} at the generalized eigenvalues
$\Lambda$ (see \cite{V7} and the footnote under that reference).

\subsection{The generic situation}

The preceding treatment does not generalize straightforwardly.
Instead, the mere division of the functional relation (\ref{FFR}) at 
$\lambda=-E_k$ by its first conjugate partner leads to the pair of formulae
\beq
2 \arg  D^{[1]\pm}(-\e^{-\mi\varphi} E_k) -\varphi \, \beta_{-1}(\vec v) =
\pi [k+\hf \pm \textstyle {N-2 \over 2(N + 2)}] \qquad
{\rm for\ } k \ {\rm even \atop odd} \ ,
\eeq
which have the outer form of Bohr--Sommerfeld quantization conditions
but are {\sl exact, albeit implicit\/}
(the left-hand sides invoke the determinants of the first
conjugate spectra, equally unknown).

However, we can write the {\sl totality\/} of such conditions
for the $L$ conjugate problems (Sec.2.1),
together with the structure equations that express
all the determinants from their spectra: namely, 
for $\ell \in {\mathbb Z} / L {\mathbb Z}$ and $k,K \ {\rm even \atop odd}$,
\bea
\label{EQC}
\textstyle \mi^{-1}
\Bigl[ \log D^{[\ell+1]\pm} (-\e^{-\mi\varphi}E_k^{[\ell]})
- \log D^{[\ell-1]\pm} (-\e^{+\mi\varphi}E_k^{[\ell]}) \Bigr]
\si-\se (-1)^\ell \varphi \beta_{-1}(\vec v) \\[4pt]
\si=\se
\pi [ k+\hf \pm \textstyle {N-2 \over 2(N+2)} ] \, , \nonumber
\eea
\bea
\label{CSE}
\log D^{[\ell]\pm} (\lambda) \si\equiv\se \lim_{K \to +\infty} \Biggl \{
\sum_ {k<K} \log (E_k^{[\ell]} + \lambda)
+ \hf \log (E_K^{[\ell]} + \lambda) \Biggr . \\[-10pt]
&& \qquad \qquad \qquad \qquad \qquad - \hf \sum_{ \{\alpha >0 \} } \Biggl.
b_\alpha^{[\ell]} ( E_K^{[\ell]} ) ^\alpha
[ \log E_K^{[\ell]} - \textstyle {1 \over \alpha} ] \Biggr \} . \nonumber
\eea
As a result, the exact spectra now appear as solutions of nonlinear,
parity-split, selfconsistent or {\sl fixed-point\/} equations of the form
\beq
\label{FPE}
{\mathcal M}^+ \{ {\mathcal E}^{\bullet +} \}= {\mathcal E}^{\bullet +}, 
\quad resp.\quad
{\mathcal M}^- \{ {\mathcal E}^{\bullet -} \}= {\mathcal E}^{\bullet -},
\eeq
where ${\mathcal E}^{\bullet \pm}$ denote the disjoint unions of
the conjugate spectra ${\mathcal E}^{[\ell] \pm}$.

Numerical tests, done for $N \le 6$ and not too big coefficients $\vec v$,
suggest that the fixed-point equations (\ref{EQC})--(\ref{CSE}) 
can be cast into {\sl contractive\/} form.
Assuming this conjecture, the problems (\ref{FPE})
get solved simply by forward iterations, which induce convergence 
toward the exact spectra starting from trial spectra 
that need only be {\sl semiclassically correct\/} (to a certain order).

\subsection{Example: homogeneous potentials $q^N$ \cite{V2}--\cite{V4}}

This is the simplest case 
(all conjugate spectra coincide, and $\lambda$ is the only variable), 
and specially so if $N>2$ (see Sec.4.1 for $N=2$, \cite{V4} for $N=1$).

\begin{figure}
{\hfill
\epsfysize=6.5cm
\epsfbox{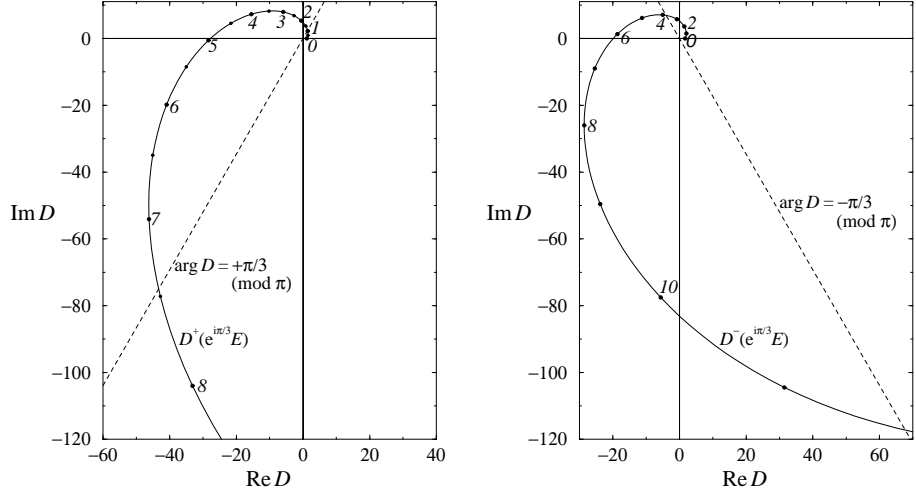}
\hfill}
\caption{\small The exact quantization formulae (\ref{HQC}) 
($+$, left; $-$, right) for the homogeneous quartic potential $q^4$: 
the curves $ \{ D^\pm(\e^{\mi\pi/3}E) \}_{E>0}$ 
(solid lines labeled by $E$-values) cross 
the (dashed) lines of prescribed argument in the complex $D$-plane.
}
\end{figure}

\begin{figure}
{\hfill
\epsfysize=8cm
\epsfbox{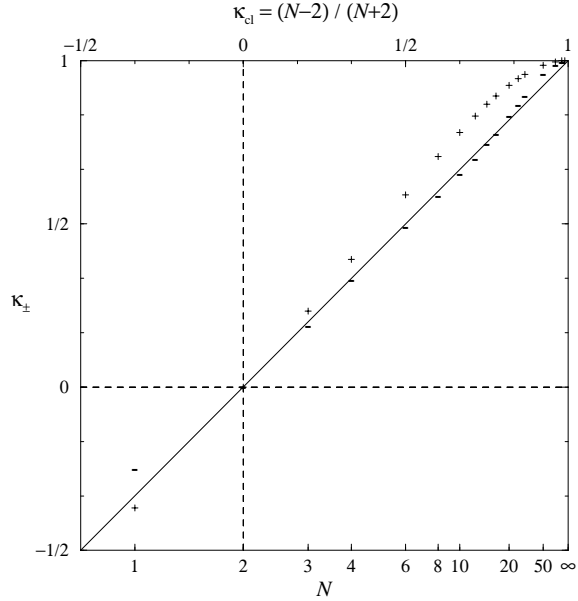}
\hfill}
\caption{\small Numerically measured linear contraction factors $\kappa_\pm$
at the fixed points (i.e., the exact $\pm$ spectra), upon iterations of the
exact quantization conditions (\ref{HQC})--(\ref{HSE}) for homogeneous
potentials $|q|^N$. As horizontal scale we use 
the value taken by $\kappa_\pm$ classically 
(when $D^\pm$ is replaced by $D_{\rm cl}^\pm$ as in eq.(\ref{CPS})):
$\kappa_{\rm cl} (N) \equiv {N-2 \over N+2}$.
}
\end{figure}

For $N>2$, the exact quantization conditions (\ref{EQC}) reduce to a single pair
(e.g., see fig.1 for $N=4$),
\beq
\label{HQC}
2 \arg  D^\pm(-\e^{-\mi\varphi} E_k) =
\pi [ k + \hf \pm \textstyle {N-2 \over 2(N \!+ 2)} ] \quad\mbox{for }
k = \textstyle{ 0,2,4,\ldots \atop 1,3,5,\ldots}
\eeq
while the structure equations (\ref{CSE}) simplify as
\beq
\label{HSE}
\arg D^\pm (-\e^{-\mi\varphi} E) \equiv
\sum _{k \ {\rm even \atop odd}}
\! \arg(E_{k}- \e^{-\mi\varphi} E) \qquad (E>0) ,
\eeq
and the asymptotic boundary condition is 
$b_\mu {E_k}^\mu  \sim k + \hf \quad {\rm for\ } k \to +\infty$.

The latter fixed-point equations have now been proven
{\sl globally contractive\/} on suitable sequence spaces \cite{A},
by methods specific to the homogeneous case (see fig.2).
(For the more general equations (\ref{EQC})--(\ref{CSE}),
contractivity remains a conjecture.)

Those very same equations (\ref{HQC})--(\ref{HSE}) have also been found 
to yield the ground states of some exactly solvable models in 2D statistical 
mechanics \cite{DT,SU} and in conformal quantum field theories \cite{BLZ} 
({\sl Bethe Ansatz\/} equations). 
A deep explanation for this coincidence of equations 
(and, by uniqueness, of their solutions)
in so different dynamical settings is not yet available.
This ``ODE/IM correspondence" (as a dictionary between
Schr\"odinger Ordinary Differential Equations and Integrable Models)
currently applies to homogeneous problems (even singular ones)
plus the supersymmetric equations of Sec.4.1 on the Schr\"odinger side,
and only to the ground (non-excited) states in integrable models.
There is now no integrable-model realization of the general 
inhomogeneous quantization conditions (\ref{EQC})--(\ref{CSE});
by extension, we still consider these as Bethe Ansatz equations,
and the general 1D Schr\"odinger equation (\ref{SSE}) as
exactly solvable through them.

\section{Extensions: realized vs desirable}

\subsection{Solving for the unknown functions $\psi$ \cite{V5}--\cite{V8}}

We keep focusing on the recessive solution $\psi_\lambda$
of eq.(\ref{SSE}) as an example.
By letting now the reference half-line vary as $[Q,+\infty)$,
we obtain parametric translates of the basic identities (\ref{BID}), essentially
$\psi_\lambda (Q) \equiv D_Q^-(\lambda ), \quad
\psi'_\lambda (Q) \equiv - D_Q^+(\lambda )$.
The computation of $\psi_\lambda (Q)$ thus amounts to 
a parametric spectral problem, and this is now exactly solvable through
parametric fixed-point equations like (\ref{FPE}), e.g.,
${\mathcal M}_Q^- \{ {\mathcal E}_Q^{\bullet -} \} 
= {\mathcal E}_Q^{\bullet-} $ for $\psi_\lambda (Q)$.
(This has also undergone successful numerical tests.)

\subsection{Toward quantum perturbation theory \cite{BW1,V9}}

By rescalings {\sl \`a la\/} Symanzik, the perturbative regime 
for an anharmonic potential like $V(q)=q^2+gq^4$ amounts to 
the $v\to +\infty$ regime for the potential $q^4+vq^2$.
This limit is not manifest, and is actually singular,
in our exact quantization formalism.
We recently obtained explicit large-$v$ behaviors 
for such spectral determinants, basically as
\beq
{ \det^\pm ( -{\d^2 \over \d q^2} + q^N + vq^M + \lambda ) \over
\det^\pm ( -{\d^2 \over \d q^2} + vq^M + \lambda ) } \sim 
{ \det_{\rm cl}^\pm ( -{\d^2 \over \d q^2} + q^N + vq^M + \lambda ) \over
\det_{\rm cl}^\pm ( -{\d^2 \over \d q^2} + vq^M + \lambda ) } 
\eeq
(for $v \to +\infty$, with $N>M$); 
i.e., the singular (divergent) large-$v$ behavior is entirely 
concentrated in the {\sl classical parts\/} of the determinants; 
these now express in terms of improper action integrals (cf. Sec.3.1),
\beq
{ \det_{\rm cl}^\pm ( -{\d^2 \over \d q^2} + q^N + vq^M + \lambda ) \over
\det_{\rm cl}^\pm ( -{\d^2 \over \d q^2} + vq^M + \lambda ) } \equiv
{ \exp \int_0^{+\infty} (q^N + vq^M + \lambda)^{1/2} \,\d q \over
\exp \int_0^{+\infty} (vq^M + \lambda)^{1/2} \,\d q } \, , \nonumber
\eeq
and the latter can be estimated in the end \cite{V9}.

\subsection{Generalizations?}

It is most desirable to push exact solvability beyond polynomial potentials.
According to some exact results, rational potentials \cite{AKT2,BLZ}
and trigonometric polynomials \cite{ZJ,D} appear as realistic targets,
as well as some differential equations of order $>2$ \cite{AKT,SUN,DDT,AKKT}.
The general Heun class is also a possible direction \cite{K},
which includes other important equations (Mathieu, Lam\'e,~\ldots).

Of course, an even greater challenge is to tackle higher-dimensional
Schr\"o\-dinger-type problems, 
in line with the original Balian--Bloch surmise itself \cite{BB}.
Exact asymptotic methods are still in the exploratory stage there, 
see for instance \cite{CV,P,BH,SI,OSIT}.

Therefore the subject is far from being closed!


\begin{thebibliography}{99}

\bibitem{AKT} T. Aoki, T. Kawai and Y. Takei, {\it New turning points in
the exact WKB analysis for higher-order ordinary differential equations\/}, 
in \cite{BM}, vol.~I, pp. 69--84.

\bibitem{AKT2} T. Aoki, T. Kawai and Y. Takei, 
{\it Algebraic analysis of singular perturbations - on exact WKB analysis\/}, 
in Japanese: S\=ugaku {\bf 45} (1993) 299--315, 
in English: Sugaku Expositions {\bf 8} (1995) 217--240.

\bibitem{AKKT} T. Aoki, T. Kawai, T. Koike and Y. Takei, {\it On the exact WKB
analysis of operators admitting infinitely many phases\/},
Adv. Math. {\bf 181} (2004) 165--189.

\bibitem{A} A. Avila, {\it Convergence of an exact quantization scheme\/}, 
Commun. Math. Phys. {\bf 249} (2004) 305--318.

\bibitem{BB} R. Balian and C. Bloch,
{\it Solutions of the Schr\"odinger equation in terms of classical paths\/},
Ann. Phys. (NY) {\bf 85} (1974) 514--545.

\bibitem{BPV} R. Balian, G. Parisi and A. Voros, {\it Quartic oscillator\/}, 
in: {\it Feynman path integrals\/} (Proceedings, Marseille 1978), 
eds. S. Albeverio {\it et al.\/}, 
Lect. Notes in Phys. {\bf 106}, Springer-Verlag, Berlin (1979) 337--360.

\bibitem{BLZ} V.V. Bazhanov, S.L. Lukyanov and A.B. Zamolodchikov,
{\it Spectral determinants for the Schr\"odinger equation and Q-operators of
conformal field theory\/}, J. Stat. Phys. {\bf 102} (2001) 567--576.

\bibitem{BW1} C.M. Bender and T.T. Wu, {\it Anharmonic oscillator\/}, 
Phys. Rev. {\bf 184} (1969) 1231--1260.

\bibitem{BW2} C.M. Bender and T.T. Wu, {\it Anharmonic oscillator II. 
A study of perturbation theory in large order\/}, Phys. Rev. {\bf D7} (1973) 
1620--1636.

\bibitem{BH} 
M.V. Berry and C.J. Howls, {\it High orders of the Weyl expansion for quantum 
billiards: resurgence of periodic orbits, and the Stokes phenomenon\/},
Proc. Roy. Soc. Lond. {\bf A447} (1994) 527--555.

\bibitem{BMK} L. Boutet de Monvel and P. Kr\'ee, 
{\it Pseudo-differential operators and Gevrey classes\/}, 
Ann. Inst. Fourier, Grenoble {\bf 17} (1967) 295--323.

\bibitem{BM} L. Boutet de Monvel ed.,
{\it Analyse alg\'ebrique des perturbations singuli\`eres\/}
(Proceedings, CIRM, Marseille--Luminy 1991), Hermann, Paris (1994).

\bibitem{BI} J. Bros and D. Iagolnitzer, {\it Causality and local analyticity:
mathematical study\/}, Ann. Inst. H. Poincar\'e {\bf A 18} (1973) 147--184
and refs. therein.

\bibitem{CV} P. Cartier and A. Voros, {\it Une nouvelle interpr\'etation 
de la formule des traces de Selberg\/}, C.R. Acad. Sci. (Paris) {\bf 307}, 
S\'erie I (1988) 143--148, and in: {\it The Grothendieck Festschrift\/} 
(vol.~II), eds. P. Cartier {\it et al.\/}, Progress in Mathematics, 
Birkh\"auser, Boston (1990), pp. 1--67.

\bibitem{C} J. Chazarain, 
{\it Formule de Poisson pour les vari\'et\'es riemanniennes\/},
Invent. Math. {\bf 24} (1974) 65--82.

\bibitem{D} \'E. Delabaere, {\it Spectre de l'op\'erateur de Schr\"odinger 
stationnaire unidimensionnel \`a potentiel polyn\^ome trigonom\'etrique\/},
C.R. Acad. Sci., S\'erie I, {\bf 314} (1992) 807--810.

\bibitem{DDP} \'E. Delabaere, H. Dillinger and F. Pham,  
{\it Exact semiclassical expansions for one-dimensional quantum oscillators\/},
J. Math. Phys. {\bf 38} (1997) 6126--6184 and refs. therein.

\bibitem{DP} \'E. Delabaere and F. Pham,  
{\it Resurgent methods in semiclassical analysis\/},
Ann. Inst. H. Poincar\'e (Physique Th\'eorique) {\bf 71} (1999) 1--94
and refs. therein.

\bibitem{DP2} \'E. Delabaere and F. Pham, 
{\it Unfolding the quartic oscillator\/}, Ann. Phys. {\bf 261} (1997) 180--218.

\bibitem{DI} R.B. Dingle, {\it Asymptotic expansions: their derivation and
interpretation\/}, Academic Press, New York (1973).

\bibitem{DT} P. Dorey and R. Tateo, 
{\it Anharmonic oscillators, the thermodynamic Bethe Ansatz 
and nonlinear integral equations\/}, J. Phys. {\bf A32} (1999) L419--L425,
and {\it On the relation between Stokes
multipliers and the T-Q systems of conformal field theory\/},
Nucl. Phys. {\bf B563} (1999) 573--602 and refs. therein
(Erratum: Nucl. Phys. {\bf B603} (2001) 581).

\bibitem{DDT} P. Dorey, C. Dunning and R. Tateo, 
{\it Differential equations for general $SU(n)$ Bethe Ansatz systems\/},
J. Phys. {\bf A 33} (2000) 8427--8442.

\bibitem{DG} J.J. Duistermaat and V.W. Guillemin, 
{\it The spectrum of positive elliptic operators and periodic bicharacteristics\/}, 
Invent. Math. {\bf 29} (1975) 39--79.

\bibitem{E} J. \'Ecalle, {\it Les fonctions r\'esurgentes\/} (vol.~1), 
Publ. Math. Univ. Paris-Sud (Orsay) 81-05 (1981) [unpublished], 
and {\it Cinq applications des fonctions r\'esurgentes\/}
(chap.1), Math. preprint 84T62, Univ. Paris-Sud, Orsay (1984) [unpublished], 
and {\it Weighted products and parametric resurgence\/}, in \cite{BM}, 
vol.~I, pp. 7--49.

\bibitem{H} E.M. Harrell, 
{\it On the rate of asymptotic eigenvalue degeneracy\/}, Commun. Math. Phys.
{\bf 60} (1978) 73--95.

\bibitem{H1} L. H\"ormander, {\it Fourier integral operators I\/}, 
Acta Math. {\bf 127} (1971) 79--183.

\bibitem{H2} L. H\"ormander, {\it The analysis of linear partial differential 
operators\/} vol.~I, Springer-Verlag, Berlin (1983) and refs. therein.

\bibitem{HKT} C. Howls, T. Kawai and Y. Takei eds., 
{\it Toward the exact WKB analysis of differential equations, 
linear or non-linear\/} (Proceedings, RIMS, Kyoto 1998), Kyoto Univ. Press, 
Kyoto (2000), and refs. therein.

\bibitem{JL} J. Jorgenson and S. Lang,
{\it Basic analysis of regularized series and products\/},
Lect. Notes in Math. {\bf 1564}, Springer-Verlag, Berlin (1993).

\bibitem{KT} T. Kawai and Y. Takei,
{\it Secular equations through the exact WKB analysis\/},
in \cite{BM}, vol.~I, pp. 85--102 and refs. therein.

\bibitem{K} T. Koike, {\it Asymptotics of the spectrum of Heun's equation and
the exact WKB analysis\/}, in \cite{HKT}, pp. 55--70.

\bibitem{L} J. Leray, {\it Probl\`eme de Cauchy I\/}, Bull. Soc. Math. Fr. 
{\bf 85} (1957) 389--429.

\bibitem{M} B. Malgrange, {\it M\'ethode de la phase stationnaire et sommation
de Borel\/}, in: 
{\it Complex analysis, microlocal calculus and relativistic quantum theory\/}
(Proceedings, Les Houches 1979), ed. D. Iagolnitzer,
Lect. Notes in Phys. {\bf 126}, Springer-Verlag, Berlin (1980) pp. 170--177.

\bibitem{OSIT} T. Onishi, A. Shudo, K.S. Ikeda, and K. Takahashi,
{\it Semiclassical study on tunneling processes via complex-domain chaos\/},
Phys. Rev. {\bf E 68} (2003) 056211 and refs. therein.

\bibitem{P} F. Pham, {\it Principe de Huygens et trajectoires complexes 
ou Balian et Bloch vingt ans apr\`es\/} (Proceedings, Grenoble 1993),
Ann. Inst. Fourier, Grenoble {\bf 43} (1993) 1485--1508.

\bibitem{SKK} M. Sato, T. Kawai and M. Kashiwara,
{\it Microfunctions and pseudo-differential equations\/}, in:
{\it Hyperfunctions and pseudo-differential equations\/} (Proceedings, Katata
1971), ed. H. Komatsu, Lect. Notes in Math. {\bf 287}, Springer-Verlag, 
Berlin (1973) pp. 265--529, and refs. therein.

\bibitem{SI} A. Shudo and K.S. Ikeda, 
{\it Complex classical trajectories and chaotic tunneling\/}, 
Phys. Rev. Lett. {\bf 74} (1995) 682--685, and 
{\it Stokes phenomenon in chaotic systems: pruning trees of complex paths with principle of exponential dominance\/}, 
Phys. Rev. Lett. {\bf 76} (1996) 4151--4154.

\bibitem{S} Y. Sibuya, {\it Global theory of a second order linear ordinary
differential operator with a polynomial coefficient\/},
North-Holland, Amsterdam (1975), and refs. therein.

\bibitem{SJ} J. Sj\"ostrand, {\it Singularit\'es analytiques microlocales\/},
Ast\'erisque {\bf 95} (1982) 1--166 and refs. therein.

\bibitem{SU} J. Suzuki, {\it Functional relations in Stokes multipliers --- 
fun with $x^6 + \alpha x^2$ potential\/},
J. Stat. Phys. {\bf 102} (2001) 1029--1047 and refs. therein.

\bibitem{SUN} J. Suzuki, {\it Functional relations in Stokes multipliers
and solvable models related to $U_q(A_n^{(1)})$}, 
J. Phys. {\bf A 33} (2000) 3507--3521.

\bibitem{V1} A. Voros, {\it Semiclassical correspondence and exact results:
the case of the spectra of homogeneous Schr\"odinger operators\/}, 
in French: C.R. Acad. Sci., S\'erie I, {\bf 293} (1981) 709--712, 
in English: J. Physique Lett. {\bf 43} (1982) L1--L4; and
{\it The return of the quartic oscillator. The complex WKB method\/},
Ann. Inst. H. Poincar\'e {\bf A 39} (1983) 211--338.

\bibitem{VC} A. Voros, {\it Spectral functions, special functions 
and the Selberg zeta function\/}, 
Commun. Math. Phys. {\bf 110} (1987) 439--465.\footnote
{In \cite{VC}, eq.(6.25) should have read
$\zeta'(-1)={\zeta'(2) \over 2\pi^2}+{1 \over 12}(1-\gamma-\log \, 2\pi)$ 
(with no consequence elsewhere).}

\bibitem{V2} A. Voros,
{\it Exact quantization condition for anharmonic oscillators
(in one dimension)\/}, J. Phys. {\bf A 27} (1994) 4653--4661.

\bibitem{V3} A. Voros,
{\it Exact anharmonic quantization condition (in one dimension)\/},
in: {\it Quasiclassical methods\/} (Proceedings, IMA, Minneapolis 1995),
eds. J. Rauch and B. Simon, IMA series vol. {\bf 95},
Springer-Verlag, New York (1997) 189--224.

\bibitem{V4} A. Voros,
{\it Airy function (exact WKB results for potentials of odd degree)\/},
J. Phys. {\bf A32} (1999) 1301--1311.\footnote
{Misprints in \cite{V4}: in eqs.(18), $D^\pm(\e^{-\mi\varphi} \lambda)$
should read $D^\pm(-\e^{-\mi\varphi} \lambda)$ (twice) and just underneath,
$[0, \e^{-\mi\varphi} \infty)$ should read $[0, -\e^{-\mi\varphi} \infty)$.}

\bibitem{V5} A. Voros, {\it Exact resolution method for general 1D
polynomial Schr\"o\-dinger equation\/}, J. Phys. {\bf A32} (1999) 5993--6007
(Corrigendum: \cite{VE}).

\bibitem{V6} A. Voros, {\it Exact quantization method 
for the 1D polynomial Schr\"odinger equation\/}\footnote{The Corrigendum 
\cite{VE} to \cite{V5} is also required for \cite{V6}, see end of Sec.3.1 
above.}, in \cite{HKT}, pp. 97--108.

\bibitem{VE} A. Voros, Corrigendum to \cite{V5}, 
J. Phys. {\bf A33} (2000) 5783--5784.$^3$

\bibitem{V7} A. Voros, {\it Exercises in exact quantization\/},
J. Phys. {\bf A33} (2000) 7423--7450, and refs. therein.\footnote
{Corrections for \cite{V7}: 1) the term ``quasi-exactly solvable" was wrongly 
put for ``supersymmetric" throughout (without practical consequence;
the two notions happen to agree when $N=2$ and 6, but not beyond);
2) regarding the Airy zeros in Table 1, $ Z_1^+\,'(0) \approx 0.0861126 $, 
$\e^{-Z_1^+\,'(0)}\approx 0.9174909 $, $\e^{-Z_1^-\,'(0)}\approx 1.2585417 $.}

\bibitem{V8} A. Voros, 
{\it ``Exact WKB integration" of polynomial 1D Schr\"odinger
(or Sturm--Liouville) problem\/}, in: {\it Differential equations and the 
Stokes phenomenon\/} (Proceedings, Groningen 2001), 
eds. B.L.J. Braaksma {\it et al.\/},
World Scientific, Singapore (2002), pp. 293--308.

\bibitem{V9} A. Voros, {\it From exact-WKB towards singular quantum 
perturbation theory\/}, Publ. RIMS, Kyoto Univ. {\bf 40} (2004) 973--990.

\bibitem{ZJ} J. Zinn-Justin, {\it Instantons in quantum mechanics: 
Numerical evidence for a conjecture\/}, J. Math. Phys. {\bf 25} (1984) 549--555,
and {\it From instantons to exact results\/}, in \cite{BM}, vol.~I, 51--68.

\end{thebibliography}
\end{document}